
\input harvmac.tex
\Title{CTP/TAMU-79/92}{{Geodesic Scattering of Solitonic Strings}
\footnote{$^\dagger$}{Work supported in part by NSF grant PHY-9106593.}}

\centerline{
Ramzi~R.~ Khuri\footnote{$^*$}{Supported by a World Laboratory Fellowship.}}
\bigskip\centerline{Center for Theoretical Physics}
\centerline{Texas A\&M University}\centerline{College Station, TX 77843}

\vskip .3in
We compute the metric on moduli space for the Dabholkar-Harvey string
soliton in $D=4$ to lowest nontrivial order in the string tension. The
metric is found to be flat, which implies  trivial scattering
of the solitons. This result is consistent with an earlier test-string
calculation of the leading order dynamical force and a computation of the
Veneziano amplitude for the scattering of macroscopic strings.

\Date{12/92}

\lref\quartet{D.~J.~Gross,
J.~A.~Harvey, E.~J.~Martinec and R.~Rohm, Nucl. Phys. {\bf B256} (1985) 253.}

\lref\dine{M.~Dine, Lectures delivered at
TASI 1988, Brown University (1988) 653.}

\lref\bpst{A.~A.~Belavin, A.~M.~Polyakov, A.~S.~Schwartz and Yu.~S.~Tyupkin,
Phys. Lett. {\bf B59} (1975) 85.}

\lref\rkinst{R.~R.~Khuri, Phys. Lett. {\bf B259} (1991) 261.}

\lref\ccrk{C.~G.~Callan and R.~R.~Khuri, Phys. Lett. {\bf B261} (1991) 363.}

\lref\rkmant{R.~R.~Khuri, Nucl. Phys. {\bf B376} (1992) 350.}

\lref\rkthes{R.~R.~Khuri, {\it Solitons and Instantons in String Theory},
 Princeton University Doctoral Thesis, August 1991.}

\lref\mono{R.~R.~Khuri, Phys. Lett. {\bf B294} (1992) 325.}

\lref\monscat{R.~R.~Khuri, Phys. Lett. {\bf B294} (1992) 331.}

\lref\rkgeo{R.~R.~Khuri, ``Geodesic Scattering of Solitonic Strings",
Texas A\&M preprint CTP/TAMU-79/92.}

\lref\dghrr{A.~Dabholkar, G.~Gibbons, J.~A.~Harvey and F.~Ruiz Ruiz,
Nucl. Phys. {\bf B340} (1990) 33.}

\lref\dabhar{A.~Dabholkar and J.~A.~Harvey,
Phys. Rev. Lett. {\bf 63} (1989) 478.}

\lref\prso{M.~K.~Prasad and C.~M.~Sommerfield, Phys. Rev. Lett. {\bf 35}
(1975) 760.}

\lref\mantone{N.~S.~Manton, Nucl. Phys. {\bf B126} (1977) 525.}

\lref\manttwo{N.~S.~Manton, Phys. Lett. {\bf B110} (1982) 54.}

\lref\mantthree{N.~S.~Manton, Phys. Lett. {\bf B154} (1985) 397.}

\lref\atiyah{M.~F.~Atiyah and N.~J.~Hitchin, Phys. Lett. {\bf A107}
(1985) 21.}

\lref\atiyahbook{M.~F.~Atiyah and N.~J.~Hitchin, {\it The Geometry and
Dynamics of Magnetic Monopoles}, Princeton University Press, 1988.}

\lref\gsw{M.~B.~Green, J.~H.~Schwartz and E.~Witten,
{\it Superstring Theory} vol. 1, Cambridge University Press (1987).}

\lref\polch{J.~Polchinski, Phys. Lett. {\bf B209} (1988) 252.}

\lref\dfluone{M.~J.~Duff and J.~X.~Lu, Nucl. Phys. {\bf B354} (1991) 141.}

\lref\dflutwo{M.~J.~Duff and J.~X.~Lu, Nucl. Phys. {\bf B354} (1991) 129.}

\lref\dfluthree{M.~J.~Duff and J.~X.~Lu, Phys. Rev. Lett. {\bf 66}
(1991) 1402.}

\lref\dflufour{M.~J.~Duff and J.~X.~Lu, Nucl. Phys. {\bf B357} (1991)
534.}

\lref\dfstel{M.~J.~Duff and K.~S.~Stelle, Phys. Lett. {\bf B253} (1991)
113.}

\lref\gh{G.~W.~Gibbons and S.~W.~Hawking, Phys. Rev. {\bf D15}
(1977) 2752.}

\lref\ghp{G.~W.~Gibbons, S.~W.~Hawking and M.~J.~Perry, Nucl. Phys. {\bf B318}
(1978) 141.}

\lref\briho{D.~Brill and G.~T.~Horowitz, Phys. Lett. {\bf B262} (1991)
437.}

\lref\raj{R.~Rajaraman, {\it Solitons and Instantons}, North Holland,
1982.}

\lref\bogo{E.~B.~Bogomolnyi, Sov. J. Nucl. Phys. {\bf 24} (1976) 449.}

\lref\dflufive{M.~J.~Duff and J.~X.~Lu, Class. Quant. Grav. {\bf 9}
(1992) 1.}

\lref\dflusix{M.~J.~Duff and J.~X.~Lu, Phys. Lett. {\bf B273} (1991)
409.}

\lref\chsone{C.~G.~Callan, J.~A.~Harvey and A.~Strominger, Nucl. Phys.
{\bf B359} (1991) 611.}

\lref\chstwo{C.~G.~Callan, J.~A.~Harvey and A.~Strominger, Nucl. Phys.
{\bf B367} (1991) 60.}

\lref\strom{A.~Strominger, Nucl. Phys. {\bf B343} (1990) 167.}

\lref\duff{M.~J.~Duff, Class. Quant. Grav. {\bf 5} (1988).}

\lref\felce{A.~G.~Felce and T.~M.~Samols, ``Low-Energy Dynamics of String
Solitons" NSF-ITP-92-155, November 1992.}

\lref\cfmp{C.~G.~Callan, D.~Friedan, E.~J.~Martinec
and M.~J.~Perry, Nucl. Phys. {\bf B262} (1985) 593.}

\lref\ckp{C.~G.~Callan,
I.~R.~Klebanov and M.~J.~Perry, Nucl. Phys. {\bf B278} (1986) 78.}

\newsec{Introduction}

In \dghrr, Dabholkar {\it et al.} constructed static multi-soliton ``cosmic
string'' solutions of the low-energy supergravity equations of the heterotic
string. The existence of this solution owes to the cancellation of the
long-range forces due to exchange of axions, gravitons and dilatons and is
reminiscent of the cancellation of gauge and Higgs forces between BPS monopoles
\mantone~. Since these static properties are also possessed by fundamental
strings  winding around an infinitely large compactified dimension, Dabholkar
{\it et al.} conjectured \dabhar~ that the soliton is actually the exterior
solution for an infinitely long fundamental string.

More recently \ccrk, the scattering of these solitons was examined using the
``test string'' approximation:
{}From the sigma model action describing the motion of a point-like
test string in a general background of axion, graviton
and dilaton fields, an effective action was derived for the motion of the
center of mass coordinate of the test string in the special background provided
by a string soliton. In addition to the zero static force, it was found that
the $O(v^2)$ velocity-dependent forces vanish as well. This result, which
implies trivial scattering of the solitons,  was found
to be consistent with expectations from the Veneziano formula.

In the low-velocity limit, multi-soliton solutions trace out geodesics in
the static solution manifold, with distance defined by the Manton metric on
moduli space manifold \manttwo. In the absence of a full time-dependent
solution to the equations of motion, these geodesics represent a good
approximation to the low-energy dynamics of the solitons. For BPS monopoles,
the Manton procedure was implemented by Atiyah and Hitchin
\refs{\atiyah,\atiyahbook}.

In this letter we compute the Manton metric on moduli space for the
scattering of the
soliton string solutions in $D=4$. We find that the metric is flat to lowest
nontrivial order in the string tension. This result implies trivial
scattering of the string solitons and is consistent with the results
of \ccrk, and thus provides even more compelling evidence for the
identification of the string soliton with the underlying fundamental string.

\newsec{Metric on Moduli Space}

We begin with a brief review of the solution of \dghrr.
The action for the massless spacetime fields
(graviton, axion and dilaton) in the presence of a source
string can be written as \dghrr
\eqn\action {S = {1\over 2\kappa^2}\int d^Dx\sqrt{g}\left(R-{1\over 2}
{{(\partial\phi)}^2}-{1\over 12}{e^{-2\alpha\phi}}H^2\right)+S_\sigma,}
with the source terms contained in the sigma model action $S_\sigma$,
\eqn\ssigma {S_\sigma=-{\mu\over 2}\int d^2\sigma(\sqrt{\gamma} \gamma^{mn}
\partial_mX^\mu\partial_nX^\nu g_{\mu\nu}e^{\alpha\phi} +
\epsilon^{mn}\partial_mX^\mu\partial_nX^\nu B_{\mu\nu}),}
with $\alpha=\sqrt{2/(D-2)}$, $H=dB$ and $\gamma_{mn}$ a worldsheet metric to
be
determined. The static solution to the equations of motion is given by \dghrr\
\eqn\ansatz{\eqalign{ds^2&=e^{A}[-dt^2+(dx^1)^2]+e^{B}d\vec x \cdot d\vec x\cr
A&={D-4\over D-2}E(r)\qquad B=-{2\over D-2}E(r)\cr
\phi&=\alpha E(r)\qquad\qquad B_{01}=-e^{E(r)},\cr}}
where $x^1$ is the direction along the string, $r=\sqrt{\vec x \cdot\vec x}$
and
\eqn\ssoln{e^{-E(r)}=\cases{1+{M\over r^{D-4}}&$D>4$\cr 1-8G\mu\ln(r)&$D=4$
\cr}}
for a single static string source. The solution can be generalized to
an arbitrary number of static string sources by linear superposition of
solutions of the ($D-2$)-dimensional Laplace's equation.
In this letter we compute the Manton metric on moduli space for the case
$D=4$, although we expect that the same result will hold for arbitrary
$D \geq 4$. Note that for $D=4$, $\phi=E$ and the metric simplifies to
\eqn\fourmet{ds^2=-dt^2+(dx^1)^2+e^{-E}d\vec x \cdot d\vec x.}

Manton's procedure may be summarized as follows. We first invert the
constraint equations of the system (Gauss' law for the case of BPS
monopoles). The corresponding time dependent field configuration does
not in general satisfy the time dependent field equations, but provides
an initial data point for the fields and their time derivatives.
Another way of saying this is that the initial motion is tangent to the
set of exact static solutions.  The resultant kinetic action obtained
by replacing the solution to the constraints in the action defines a
metric on the parameter space of static solutions. This metric defines
geodesic motion on the moduli space\manttwo.

We now assume that each string source possesses velocity
$\vec\beta_n, n=1,...,N$ in the two-dimensional transverse space $(23)$. This
will appear in the contribution
of the sigma-model source action to the equations of motion in the form
of ``moving" $\delta$-functions $\delta^{(2)}(\vec x-\vec a_n(t))$, where
$\vec a_n(t)\equiv \vec A_n + \vec\beta_n t $ (here $\vec A_n$ is the
initial position of the $n$th string source).

The equations of motion following from $S$ are complicated and nonlinear
(see \dghrr),
and it is remarkable that such a simple ansatz as above could provide a
solution in the static limit. In the time dependent case, we are even less
likely to be so fortunate. In order to make headway in solving even the
$O(\beta)$ time dependent constraints, we make the simplifying assumption
that $8G\mu << 1$ (this is equivalent to assuming that each cosmic string
produces a small deficit angle). It turns out that to linear order in $\mu$
an $O(\beta)$ solution to the constraint equations is given by
\eqn\orderbeta{\eqalign{e^{-E(\vec x,t)}&=1-8G\mu\sum_{n=1}^N{\ln(
\vec x - \vec a_n(t))}\cr g_{00}&=-g_{11}=-1,\qquad g^{00}=-g^{11}=-1\cr
g_{ij}&=e^{-E}\delta_{ij},\qquad g^{ij}=e^E\delta_{ij}\cr
g_{0i}&=8G\mu\sum_{n=1}^N{\vec\beta_n\cdot \hat x_i \ln(\vec x - \vec a_n(t))},
\qquad g^{0i}=e^Eg_{0i}\cr
H_{10j}&=\partial_j e^E\cr
H_{1ij}&=\partial_i g_{0j} - \partial_j g_{0i},\cr}}
where $i,j=2,3$.

The kinetic Lagrangian is obtained by replacing the expressions for the
fields in \orderbeta\ in $S$. Since \orderbeta\ is a solution to
order $\beta$, the leading order terms in the action (after the
quasi-static part) is of order $\beta^2$. The source part of the action
($S_2$) now represents moving string sources, and its only contribution to
the kinetic Lagrangian density is of the form $(\mu/2)\beta_n^2$ for each
source. The nontrivial elements of the metric on moduli space must therefore
be read off from the $O(\beta^2)$ part of the massless fields effective action.
In principle one must add a Gibbons-Hawking surface term (GHST)
in order to cancel the double derivative terms in $S$ (see
\refs{\gh\ghp\briho\rkmant{--}\monscat}). In this case, however, the GHST
vanishes to $O(\beta^2)$. To lowest nontrivial order in $\mu$, the kinetic
Lagrangian density is computed to be
\eqn\lkin{{\cal L}_{kin}={1\over 2\kappa^2}\left(2\dot E^2-(\partial_m
g_{0k})^2\right).}
Henceforth we simplify to the case of two strings with velocities
$\vec\beta_1$ and $\vec\beta_2$ and positions $\vec a_1$ and $\vec a_2$.
Let $\vec X_n\equiv \vec x - \vec a_n, n=1,2$.
Our moduli space consists of the configuration space of the relative
separation vector $\vec a\equiv \vec a_2 - \vec a_1$. We now compute the
metric on moduli space by integrating \lkin\ over the $(23)$ space.
It turns out that the self-terms vanish on integration over the
two-space. We are then left with the interaction terms, which may be
written as
\eqn\lint{{\cal L}_{int}={64G^2\mu^2\over \kappa^2}
\left[{2(\vec\beta_1 \cdot \vec X_1)(\vec\beta_2 \cdot \vec X_2)\over
X_1^2 X_2^2} -{(\vec\beta_1 \cdot \vec\beta_2)(\vec X_1\cdot \vec X_2)\over
X_1^2 X_2^2}\right].}
The most general answer obtained by integrating \lint\
over the transverse two-space is of the form
\eqn\intform{L_{int}(\vec a)=2f(a)\vec\beta_1 \cdot \vec\beta_2 +
2g(a)(\vec\beta_1\cdot\hat a)(\vec\beta_2\cdot\hat a).}
We compute $f$ and $g$ by integrating \lint\ for only two
configurations. In both cases, $\vec\beta_1$ is parallel to $\vec\beta_2$. The
first case has the velocities parallel to $\vec a$ and yields
\eqn\para{L_{int}(a)=(2f+2g)\beta_1\beta_2}
while the second case has the velocities perpendicular to $\vec a$ and
yields
\eqn\perp{L_{int}(a)=2f\beta_1\beta_2.}
In this way we can compute both $f$ and $g$.
A slightly tedious but straightforward computation yields
\eqn\fgans{g=-2f=-{64g^2\mu^2\pi\over \kappa^2}\left(1-{\ln 2\over
2}\right),}
and thus all the metric elements are constant. In two-dimensions, this implies
that the metric on moduli space is flat (being of the form
$dr^2+Ar^2d\theta^2$, where $A$ is a constant), and therefore has
straight-line geodesics in the static solution manifold. To this
approximation, then, the low-energy scattering is trivial, i.e. the strings
do not deviate from their initial trajectories.

\newsec{Discussion}

The above result is in perfect agreement with those of \ccrk. A flat metric
was also found \monscat\ for the recently constructed
heterotic multimonopole solution of \mono.
It should be noted that in \ccrk\ the ``test-fivebrane" approach also yielded
a zero dynamical force, and consequently trivial scattering, in the low-energy
limit for the class of string solitons with fivebrane
structure\refs{\dfluone\dflutwo\strom\chsone{--}\chstwo}.
A different result, however, was found in
\rkmant\ for the metric on moduli space for the scattering of heterotic
fivebranes. In that case, however, the self-terms were found to be divergent
and we were compelled to extract information from the convergent interaction
terms in what amounted to a regularization procedure. This difficulty with the
regularization is probably the cause of the
apparent contradiction between the two results for fivebrane scattering.
Fortunately, for both the string and monopole solutions, we do not encounter
any divergence in the self-terms and do not have to resort to
regularization. In both cases, we can explicitly compute the metric on
moduli space, which turns out to be flat, in direct agreement with
test-string and test-monopole calculations.

More recently in \felce, a flat metric was also found for the fivebranes
using a different gauge choice than in \rkmant.
For the fivebrane analogue of the
the Veneziano amplitude calculation of \ccrk\ (which yielded trivial
scattering for macroscopic winding state strings), we must await the
construction of the conjectured dual
theory of fundamental fivebranes\refs{\duff,\strom,\dfluone,\dflutwo}.

\vfil\eject
\listrefs
\bye